\documentclass[conference]{IEEEtran}
\usepackage{amsmath,amsfonts,amsthm}
\usepackage{algorithmic}
\usepackage{algorithm}
\usepackage{array}
\usepackage[caption=false,font=normalsize,labelfont=sf,textfont=sf]{subfig}
\usepackage{textcomp}
\usepackage{stfloats}
\usepackage{url}
\usepackage{verbatim}
\usepackage{graphicx}
\usepackage{cite}
\usepackage{hyperref}
\hypersetup{colorlinks=true,allcolors=blue}
\usepackage{hypcap}
\usepackage{amssymb}
\usepackage{psfrag}
\usepackage{siunitx}

\IEEEoverridecommandlockouts
\usepackage{cite}
\usepackage{amsmath,amssymb,amsfonts}
\usepackage{algorithmic}
\usepackage{graphicx}
\usepackage{textcomp}
\usepackage{xcolor}
\def\BibTeX{{\rm B\kern-.05em{\sc i\kern-.025em b}\kern-.08em
    T\kern-.1667em\lower.7ex\hbox{E}\kern-.125emX}}

\makeatletter
\def\ps@IEEEtitlepagestyle{%
  \def\@oddfoot{\mycopyrightnotice}%
  \def\@oddhead{\hbox{}\@IEEEheaderstyle\leftmark\hfil\thepage}\relax
  \def\@evenhead{\@IEEEheaderstyle\thepage\hfil\leftmark\hbox{}}\relax
  \def\@evenfoot{}%
}
\def\mycopyrightnotice{%
  \begin{minipage}{\textwidth}
  \centering \scriptsize
  Copyright~\copyright~2024 IEEE. Personal use of this material is permitted. Permission from IEEE must be obtained for all other uses, in any current or future media, including\\reprinting/republishing this material for advertising or promotional purposes, creating new collective works, for resale or redistribution to servers or lists, or reuse of any copyrighted component of this work in other works by sending a request to pubs-permissions@ieee.org.
  \end{minipage}
}
\makeatother
\begin{document}

\title{Modeling and Control of a Novel Bi-Quadcopter with Auxiliary Thruster Mechanism\\
\thanks{This work was supported by IIT Palakkad Technology IHub Foundation under the grant: IPTIF/TD/IP/007.}
}

\author{\IEEEauthorblockN{ Vijay Reddy Vundela}
\IEEEauthorblockA{\textit{Electrical Engineering} \\
\textit{Indian Institute of Technology Palakkad}\\
Kerala, India \\
vijaygiri502@gmail.com}
\and
\IEEEauthorblockN{ Vijay Muralidharan}
\IEEEauthorblockA{\textit{Assistant Professor} \\
\textit{Electrical Engineering} \\
\textit{Indian Institute of Technology Palakkad}\\
Kerala, India \\
vijay@iitpkd.ac.in}
}

\maketitle
\begin{abstract}
In this paper, a new under-actuated Bi-Quadcopter Unmanned Aerial Vehicle is introduced. The proposed drone configuration can be controlled similar to a Bicopter. The dynamics of the proposed Bi-Quadcopter is developed using the Newton-Euler approach. Using the force decomposition technique, a mapping between the control wrench and actuator inputs is developed. A nonlinear position control is applied for the Bi-Quadcopter using the quaternion-based cascaded attitude controller. The performance of the proposed UAV with the control algorithm is verified through simulations. Finally, the actuator failure scenarios were analyzed. 
\end{abstract}


\IEEEpeerreviewmaketitle

\section{Introduction}
The demand has increased for UAVs with capabilities such as heavy payload carrying capacity, entering into narrow areas (for example, in the construction fields and disaster-prone areas) with agile maneuvers to deal with complex missions. The most common UAVs are fixed-wing drones, propeller-based drones, and hybrid UAVs.  The Propeller-based UAVs are most suitable for indoor environments. 

There are numerous types of propeller-based UAVs in the literature with different configurations(the propeller arrangement and the number of actuators decide the type of configuration). The most common conventional fixed-pitch vehicles are quadcopters\cite{mellinger2011minimum}. The quadcopter consists of four(minimal actuators required to stabilize any UAV in 3D space) propellers. As the number of propellers increases, the payload capacity increases which leads to the development of conventional hexacopter\cite{jiao2020fault} and octacopter\cite{er2013development}. 
The main drawbacks in conventional fixed-pitch propeller UAVs are, that they mainly work based on increasing and decreasing the rotor speed and performing required maneuvers, which lags the faster and more aggressive maneuvers, and also the resultant thrust axis with respect to the body frame is towards motion direction which leads to the increase in the air resistance on the propellers\cite{li2021driving}.
This leads to the introduction of tilting propellers over conventional UAVs \cite{ryll2012modeling,kamel2018voliro,kastelan2015fully}. The tilting vehicles are capable of performing aggressive maneuvers with full actuation. 
Bicopter is one of such titling UAVs with under actuation that have minimal actuators(two propellers and two tilting servos) which consume less power and perform more aggressive tracking \cite{taherinezhad2022robust,albayrak2019design,qin2020gemini}.   However, the main limitations of the Bicopter are: the total thrust required to compensate for the gravitational force is shared by only two propellers which leads to the motor operating point shifting to the saturation region and hence restricting the payload capacity and also the swing burden on the servos is very high. Also, in the case of actuator failures, the control re-allocation is not possible during drone recovery or trajectory tracking. 

\begin{figure}[t]
\centering
\includegraphics[width=3.4in]{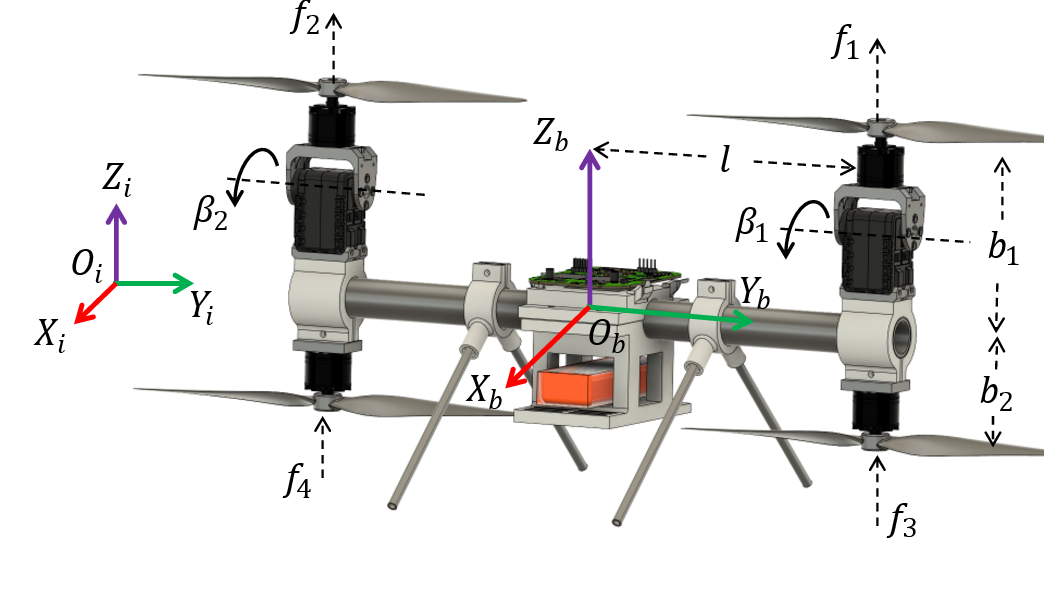}
\caption{Schematic of Bi-Quadcopter drone.}
\label{schematic}
\end{figure}

To avoid the aforementioned difficulties, in this work, two additional thrust-sharing propellers are added to the Bicopter as shown in Figure \ref{schematic}. The proposed Bi-Quadcopter mechanism is similar to a conventional Bicopter, however, it has two auxiliary propellers to share the total thrust demanded by the UAV and is hence suitable for heavy-lift applications. A comparable heavy-lift UAV found in the literature is a quadcopter with two tilting thrusters \cite{ozdougan2022design}.  The main advantages of the proposed mechanism are,
\begin{enumerate}
    \item Since the total thrust demanded is shared by the four propellers, the propeller size is reduced compared to the Bicopter, which ensures the compactness of the UAV and the burden on the servos will be reduced during the tilting.
    \item In case of actuator failures, the control re-allocation is possible due to the advantage of the presence of redundant actuators.
    \item The reaction torque due to differential thrust during Bicopter roll can be avoided due to the co-axial nature of the proposed Bi-Quadcopter.
\end{enumerate}

The main contributions of this paper are:
\begin{enumerate}
    \item Derivation of the mathematical model of the proposed heavy-lift Bi-Quadcopter with auxiliary thrusters. 
    \item Developing a control allocation method using the force decomposition approach for the proposed mechanism. Analysis of the actuator failure scenarios on the control allocation is also performed.
    \item Non-linear position control and stability verification through simulations.
\end{enumerate}

The remainder of the paper is organized as follows: Section \ref{dynamics} discusses the dynamics of the Bi-Quadcopter with auxiliary thrusters. A control allocation method using force decomposition is discussed in section \ref{allocation}. A nonlinear position and attitude control of the Bi-Quadcopter is discussed in Section \ref{control} followed by the simulations in Section \ref{simulation} and finally concluding remarks in Section \ref{conclusion}.

\section{Dynamics of Bi-Quadcopter with Auxiliary Thrusters}
\label{dynamics}
Consider the inertial frame $O_i,( X_i, Y_i, Z_i)$ with the North-West-Up (NWU) convention and the body coordinate frame $O_b, (X_b, Y_b, Z_b)$ attached to the body center of mass. The schematic of the proposed UAV with the coordinate frames is shown in Figure \ref{schematic}. Let the vectors $p = [x \ y \ z]^{\top} \in \mathbb{R}^3$ and  $v = [v_x \ v_y \ v_z]^{\top} \in \mathbb{R}^3$ be the position and velocity of the center of mass respectively expressed in the inertial frame. Also, let $R\in SO(3)$ and $\omega = [\omega_1 \ \omega_2 \ \omega_3]^{\top} \in \mathbb{R}^{3}$ be the attitude and angular velocity of the Bi-Quadcopter respectively expressed in body frame. Here, the set $SO(3)$ is the group of special orthogonal matrices.
The dynamics of the Bi-Quadcopter drone using Newton-Euler equations can be written as
\begin{equation}
\begin{gathered}
    \dot{p} = v
\\
    m\dot{v} = -mge_{3} + RF,
\end{gathered}
\label{transl}
\end{equation}
\begin{equation}
\begin{gathered}
    \dot{R} = R\widehat{\omega}
\\
    J \dot{\omega} = -\widehat{\omega} J \omega + \tau.
\end{gathered}
\label{rotat}
\end{equation}
Here, $F = [F_x \ F_y \ F_z]^{\top} \in \mathbb{R}^3$ and $\tau = [\tau_x \ \tau_y \ \tau_z]^{\top}\in \mathbb{R}^3$ are the set of external forces and torques respectively acting on the UAV body, $m$ is the mass of the Bi-Quadcopter, $g$ is the acceleration due to gravity and $J = \text{diag}([J_x \ J_y \ J_z]) \in \mathbb{R}^{3\times 3}$ is the moment-of-inertia. The notation $\widehat{\omega} \in \mathbb{R}^{3\times 3}$ in (\ref{rotat}) is the skew-symmetric matrix representation of the vector $\omega$.
Note that the vertical distance between the top and bottom propellers should be sufficiently high to avoid the airflow resistance effect of the top and bottom propellers \cite{leishman2006principles,paulos2018emulating,prothin2013vectoring,ibrahim2020rotor}.
The UAV is assumed to be controlled directly through the propeller forces and the servo's tilt angles $u = \left[f_1 \ f_2 \ f_3 \ f_4 \ \beta_1 \ \beta_2\right]^{\top} \in \mathbb{R}^{4}\times (\mathbb{S}^{1})^{2}$.

Consider, $F_i = [0 \ 0 \ f_i]^{\top} \in \mathbb{R}^{3}$, $i = 1,2,3,4$, the thrust produced by each motor in the propeller frame. The virtual forces $F$ and torques due to thrust $\tau_f \in \mathbb{R}^3$ acting on the Bi-Quadcopter UAV body frame can be expressed as follows
\begin{equation}
\begin{gathered}
 F = R_{Y}(\beta_1)F_1+ R_{Y}(\beta_2)F_2 + F_3 + F_4 \\
 \tau_f = d_{1}\times R_{Y}(\beta_1)F_1 + d_{2} \times R_{Y}(\beta_2)F_2 
   \\ + d_{3} \times F_3 + d_{4} \times F_4
\label{torque_eq}
\end{gathered}
\end{equation}

where, 
\begin{equation*}
\ d_1 = \begin{bmatrix}
 0 \\
 l\\
 b_1
\end{bmatrix},
\ d_2 = \begin{bmatrix}
 0 \\
 -l\\
 b_1
\end{bmatrix}
,
\ d_3 = \begin{bmatrix}
 0 \\
 l\\
 -b_2
\end{bmatrix}
,
\ d_4 = \begin{bmatrix}
 0 \\
 -l\\
 -b_2
\end{bmatrix}
\end{equation*}

and $R_Y(\beta_1)$, $R_Y(\beta_2) \in SO(3)$ are the rotation matrices representing the tilting of the servos around the body $Y_b$-axis and the parametrization is given below

\begin{equation}
 R_{Y}(\beta_i)=\begin{bmatrix}
  \cos(\beta_i) & 0 & \sin (\beta_i)\\
0 & 1 & 0 \\
-\sin (\beta_i) & 0 & \cos(\beta_i)
\end{bmatrix}; i = {1,2}. 
\label{rotations}
\end{equation}

Consider that propellers 1 and 4 are spinning counterclockwise (CCW) and propellers 2 and 3 are spinning in clockwise (CW) directions. The reaction torque due to the propeller spin $\tau_r$ expressed in the body frame as
\begin{equation}
    \tau_r = -R_Y(\beta_1)k_rF_1 + R_Y(\beta_2)k_rF_2 + k_rF_3 - k_rF_4.
\end{equation}
Here the coefficient $k_r$ is the ratio of torque to thrust constants of the respective propeller. Hence, the total torque acting on the UAV body is $\tau = \tau_f + \tau_r$.

\section{Control Allocation}
\label{allocation}
The external forces and torques together acting on UAV are also known as control wrench $[F^{\top}, \ \tau^{\top}]^{\top} \in \mathbb{R}^{6}$. The design of the control wrench for proper trajectory tracking is discussed in Section \ref{control}. In this Section, the control allocation relation which maps the designed control wrench with the actuator inputs $u$ is discussed.
By rewriting the external forces and torques using (\ref{torque_eq}) and (\ref{rotations}),
\begin{equation}
\begin{gathered}
F_x = f_{1}\sin(\beta_{1})+f_{2} \sin(\beta_{2}) \\
F_y = 0 \\
F_z = f_{1} \cos(\beta_{1})+f_{2} \cos(\beta_{2}) + f_{3} + f_{4} \\
\tau_x = l(f_{1} \cos( \beta_{1})-f_{2} \cos (\beta_{2})) + l (f_3 - f_4)\\ - k_r(f_1\sin(\beta_1) - f_2\sin(\beta_2)) \\
 \tau_y = b_1f_{1} \sin (\beta_{1})+b_1f_{2} \sin (\beta_{2})\\
\tau_z =  k_r(f_3 - f_4) - k_r(f_1\cos(\beta_1)  - f_2\cos(\beta_2)) \\ -l(f_{1} \sin (\beta_{1})+f_{2} \sin( \beta_{2}))
\end{gathered}
\label{torque_force_vect}
\end{equation}

By observing the above equations, the dependency in $F_x$ and $\tau_y$ can be written as $F_x = \frac{\tau_y}{b_1}$ and $F_y = 0$. Hence, the number of independent equations is four with six actuator inputs. The mapping equations above involve nonlinear and coupled input terms, and also the total number of actuators becomes six, hence the simple inverse of the control allocation matrix is not possible. To decouple the input terms, the force decomposition approach through variable transformation is used in this work\cite{kamel2018voliro,invernizzi2017geometric,li2021driving}. The non-linear allocation relation can be transferred into a linear relation by decomposing the forces generated by each propeller as vertical and horizontal forces. Consider, $F_{dec}\in \mathbb{R}^{6\times 1}$ is a force decomposition vector and is expressed as follows

\begin{equation}
F_{dec}=\left[\begin{array}{l}
F_{1V}=f_1 \cos \beta_1 \\
F_{1 L}=f_1 \sin \beta_1 \\
F_{2 V}=f_2 \cos \beta_2 \\
F_{2 L}=f_2 \sin \beta_2 \\
F_3=f_3 \\
F_4=f_4
\end{array}\right].
\end{equation}

Now, the virtual forces and torques in (\ref{torque_force_vect}) are written in terms of $F_{dec}$ as shown below.
\begin{equation}
\begin{array}{ll}
\left[\begin{array}{ll}
F_z \\
\tau
\end{array}\right] &= 
\left[\begin{array}{cccccc}
1 & 0 & 1 & 0 & 1 & 1\\
l  & -k_r & -l  & k_r & l  & -l \\
0 & b_1  & 0 & b_1  & 0 & 0\\
-k_r & -l  & k_r & l  & k_r & -k_r
\end{array}\right]
\left[\begin{array}{llllll}
F_{1V} \\
F_{1L} \\
F_{2V} \\
F_{2L} \\
F_3   \\
F_4
\end{array}\right]
\\
&= [A_{st}]F_{dec}
\label{static}
\end{array}
\end{equation}

Here, $A_{st}\in \mathbb{R}^{4\times 6}$ is a static allocation matrix with full row rank ($=4$) and the linear relation admits infinity solutions. The solution $F_{dec}$ can be obtained using the Moore-Penrose pseudo inverse of the matrix $A_{st}$ as follows.
\begin{equation}
    F_{dec} = [A^{\dag}_{st}] \left[\begin{array}{cc}
       F_z  \\
         \tau 
    \end{array}\right]
\end{equation}

Since the number of actuators is more compared to the degrees of freedom (DoF) achievable, the Moore-Penrose pseudo inverse $A^{\dag}_{st}$ is the minimum norm solution of the equation (\ref{static}) which also minimizes the two norm of the solution $F_{dec}$ such as $||F_{dec}|| = \sqrt{f^{2}_1 + f^{2}_2 + f^{2}_3 + f^{2}_4}$ that enables the optimal distribution of the thrust commands among the propellers which further minimizes the energy consumption and the overall control effort\cite{invernizzi2017geometric,kamel2018voliro}. The $A^{\dag}_{st}\in \mathbb{R}^{6\times4}$ is obtained as

\begin{equation}
A^{\dag}_{st} = 
\left[\begin{array}{cccc}
\frac{1}{4} & \frac{l(l^2+3k_r^2)}{4\,(l^2 + k_r^2)^2 } & 0 & \frac{-k_r(3l^2+k_r^2)}{4\,(l^2 + k_r^2)^2}\\
0 & \frac{-k_r^3}{2\,(l^2 + k_r^2)^2} & \frac{1}{2\,b_1 } & \frac{-l^3}{2\,(l^2 + k_r^2)^2 }\\
\frac{1}{4} & \frac{-l(l^2+3k_r^2)}{4\,(l^2 + k_r^2)^2 } & 0 & \frac{k_r(3l^2+k_r^2)}{4\,(l^2 + k_r^2)^2}\\
0 & \frac{k_r^3}{2\,(l^2 + k_r^2)^2} & \frac{1}{2\,b_1 } & \frac{l^3}{2\,(l^2 + k_r^2)^2 }\\
\frac{1}{4} & \frac{l}{4\,(l^2+k_r^2) } & 0 &\frac{k_r}{4\,(l^2+k_r^2) }\\
\frac{1}{4} & \frac{-l}{4\,(l^2+k_r^2) } & 0 &\frac{-k_r}{4\,(l^2+k_r^2) }
\end{array}\right] 
\end{equation}

Finally, the actuator inputs are obtained as follows.

\begin{equation}
\begin{aligned}
& f_1=\sqrt{F_{1 V}^2+F_{1 L}^2} \\
& f_2=\sqrt{F_{2 V}^2+F_{2 L}^2} \\
& f_3=F_3 \\
& f_4=F_4 \\
& \beta_1=\operatorname{atan2}\left(F_{1L}, F_{1 V}\right) \\
& \beta_2=\operatorname{atan2}\left(F_{2 L}, F_{2 V}\right)
\end{aligned}
\label{final_actuator}
\end{equation}

\subsection{Bi-Quadcopter propeller failure analysis}
One of the advantages of the proposed Bi-Quadcopter is actuator redundancy. Due to the presence of the additional propellers, the control re-allocation is possible when some of the actuators fail \cite{werink2019control}. In this paper mainly two possible control re-allocation cases during actuator failures are discussed.
\subsubsection*{Case 1: If any one of the bottom propellers fail}
Without loss of generality, consider thruster 4 fails (among thrusters 3 or 4). The effect of the thruster 4 on the torque $\tau$ and the force $F_z$ is zero and hence the effect of the sixth column is removed from the static allocation matrix $A_{st}$ as shown below

$
\begin{array}{ll}
\left[\begin{array}{ll}
F_z \\
\tau
\end{array}\right] &= 
\left[\begin{array}{ccccc}
1 & 0 & 1 & 0 & 1\\
l  & -k_r & -l  & k_r & l \\
0 & b_1  & 0 & b_1  & 0 \\
-k_r & -l  & k_r & l  & k_r 
\end{array}\right]
\left[\begin{array}{lllll}
F_{1V} \\
F_{1L} \\
F_{2V} \\
F_{2L} \\
F_3  
\end{array}\right]
\\
&= [A1_{st}]F1_{dec}
\label{static2}
\end{array}.
$

The solution for $F1_{dec}$ exists since the full row rank($=4$) still holds true for the matrix $A1_{st}$ and the pseudo inverse of  $A1_{st}$ can be obtained as a constant matrix. The final actuator inputs $u_1 = [f_1 \ f_2 \ f_3 \ \beta_1 \ \beta_2]^{\top}\in \mathbb{R}^5$ can be obtained using the relation in (\ref{final_actuator}).

\subsubsection*{Case 2: If both the bottom propellers fail}
This is a special case where the proposed Bi-Quadcopter works as a conventional Bicopter when both the bottom propellers fail. This case is also suitable for certain tasks where the bottom propellers can be turned off for energy saving (low payload). The control mapping after removing the effect of the bottom propellers is given below

$
\begin{array}{ll}
\left[\begin{array}{ll}
F_z \\
\tau
\end{array}\right] &= 
\left[\begin{array}{cccc}
1 & 0 & 1 & 0\\
l  & -k_r & -l  & k_r \\
0 & b_1  & 0 & b_1 \\
-k_r & -l  & k_r & l 
\end{array}\right]
\left[\begin{array}{llll}
F_{1V} \\
F_{1L} \\
F_{2V} \\
F_{2L} 
\end{array}\right]
\\
&= [A2_{st}]F2_{dec}.
\label{static2}
\end{array}
$

The $det(A2_{st}) = 4b_1(l^2 + k_r^2)$ and the solution $F2_{dec}$ is obtained using inverse of the matrix of $A2_{st}$. The actuator inputs  $u_2 = [f_1 \ f_2  \ \beta_1 \ \beta_2]^{\top}\in \mathbb{R}^4$ can be obtained by substituting $F2_{dec}$ in (\ref{final_actuator}).

\section{Position Control}
\label{control}
Let, $p_d, v_d\in \mathbb{R}^3$  be the desired position and desired velocity trajectories in 3D Euclidean space. Consider the error vectors in position and velocity 

\begin{equation}
\begin{aligned}
    p_e = p - p_d, \\ v_e = v - v_d. 
 \end{aligned}
\end{equation}
The position trajectory tracking can be achieved by controlling the magnitude of the force vector and the direction of the force vector with respect to the inertial frame. Since $F_x$ is dependent in the case of the Bi-Quadcopter, the force in vertical direction $F_z$ with respect to the body frame is the only control input for the position control, and the term $RF$ in equation (\ref{transl}) is split into $RF = RF_xe_1 + RF_ze_3$. The desired force $F_{zd} \in \mathbb{R}$ required along the body $z$-axis is obtained from the desired force vector $F_{des}\in \mathbb{R}^{3}$ as follows
\begin{equation}
    F_{zd} = (F_{des} - RF_xe_1)\cdot (Re_3).
    \label{f_z}
\end{equation}
Proposing the following PD controller for the position trajectory tracking control\cite{mellinger2011minimum}

\begin{equation}
    F_{des} = - k_p\cdot p_e - k_d \cdot v_e +mge_3 + m\dot{v}_d
    \label{force}
\end{equation}
where $k_p >0$ and $k_d >0$ are position controller parameters. From (\ref{force}), the desired direction of the third body fixed axis $Z_{bd}$ with the assumption $\|F_{des}\| \neq 0$ is taken as 
\begin{equation}
    Z_{bd} = \frac{ F_{des}}{\|F_{des}\|}.
\end{equation}
The desired attitude/rotation matrix $R_c \in SO(3)$ required for the position trajectory tracking can be computed using the vector $Z_{bd}$ and by selecting a desired heading vector $X_{bc}$  
\begin{equation}
X_{bc}  = [\cos(\psi) \  sin(\psi) \ 0]^{\top}
\end{equation}
where $\psi$ is the desired heading angle. Note that $X_{bc}\times Z_{bd} \neq 0$. Constructing a unit vector $Y_{bd}$ which is orthogonal to $X_{bc}$ and $Z_{bd}$, given by $Y_{bd} = \frac{Z_{bd}\times X_{bc}}{\|Z_{bd}\times X_{bc}\|}$. Finally, $X_{bd} = Y_{bd}\times Z_{bd}$, and the desired attitude is given by 
\begin{equation}
    R_c = \left[ X_{bd} \ Y_{bd} \ Z_{bd}\right]
\end{equation}
The attitude controller is now implemented to drive the drone attitude $R$ towards the desired attitude $R_c$.

\subsubsection*{Attitude Control} The attitude control for the Bi-Quadcopter is developed using the quaternion representation of the attitude $R$ ( for example, see \cite{fresk2013full}). Consider $q = [q_0 \ q_1 \ q_2 \ q_3]^{\top}  = [\eta \ \epsilon]^{\top} \in \mathbb{R}^4$, the quaternion representation of the attitude $R$ with $\eta$ and $\epsilon$ as the scalar and vector parts of $q$ respectively. The attitude kinematics $\dot{q}$ in terms of the quaternion variables and the quaternion attitude error variable $q_e$ are given below. 

\begin{equation}
    \dot{q} = \frac{1}{2} q\otimes\omega ; \ q_e  =q_c{ }^* \otimes q=\left[\eta_e \ \epsilon_e\right]^T 
\end{equation}

where, $q_c$ is the computed attitude $R_c$ in terms of quaternion variables.  The notation $\otimes$ is a quaternion multiplication and $\left(\cdot \right)^*$ represents the quaternion conjugate operation\cite{fresk2013full}. Consider an error function $e_{err}$ in terms of quaternion as
\begin{equation}
\begin{aligned}
e_{err} = \operatorname{sign}\left(\eta_{\mathrm{e}}\right) \frac{\varphi}{\sin \left(\frac{\varphi}{2}\right)} \boldsymbol{\epsilon}_e \\
\varphi  =2 \cdot \operatorname{atan} 2\left(\left\|\boldsymbol{\epsilon}_{\mathrm{e}}\right\|, \eta_{\mathrm{e}}\right) 
\end{aligned}
\end{equation}

\begin{figure}[b]
\centering
\includegraphics[width=3.4in]{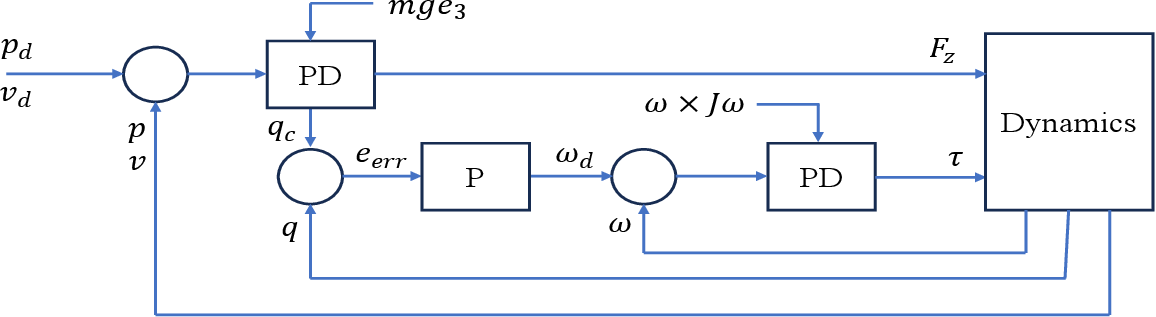}
\caption{Bi-Quadcopter position control flow diagram.}
\label{flow}
\end{figure}

In terms of angle-axis representation of the attitude, the terms $e_{err}$ and $\varphi$ are also known as the axis and the angle of the error attitude respectively\cite{zhou2017unified}. The attitude control is implemented in two stages using the cascaded approach\cite{lyu2017hierarchical}. The outer loop is an attitude-level proportional controller that gives desired angular velocity $\omega_d$ with respect to body frame based on the attitude error function $e_{err}$ as follows.

\begin{equation}
    \omega_d = -k_p^{q}\cdot e_{err}
    \label{omega_d}
\end{equation}
Here, $k_p^{q} > 0$ is the outer loop controller parameter.  Further, the obtained $\omega_d$ is passed to the inner loop. The inner angular rate-level loop is a PD controller that gives the desired torque based on the error in angular velocity $\omega_e = \omega - \omega_d$. The controller is given below.

\begin{equation}
    \tau_d = - K_p^{\omega} \cdot \omega_e - K_d^{\omega} \cdot \frac{d\omega_e}{dt} + \omega \times J \omega
    \label{torque}
\end{equation}
where $K_p^{\omega}$ and $K_d^{\omega}$ are inner loop controller parameters and are positive definite. Both desired force and torque control laws from (\ref{f_z}) and (\ref{torque}) are further substituted in equation (\ref{final_actuator}) to obtain the desired actuator inputs $u$. The Bi-Quadcopter position control flow is summarized in Figure \ref{flow}.

\begin{table}[htbp]
\caption{Bi-Quadcopter and Controller Paramters}
\begin{center}
\begin{tabular}{|c|c|c|}
\hline
\textbf{Parameter} & \textbf{Value}& \textbf{Units}
\\
\hline 
$m$ & 5 & $kg$ 
\\
\hline
 $l$ & 0.2539 & $m$ 
\\
\hline
$b_1$ & 0.14838 & $m$
\\
\hline
$g$ & 9.8 & $m \cdot s^{-2}$
\\
\hline
$J$ & \text{diag}([0.366, 0.171,  0.391])  & $kg \cdot m^2$
\\
\hline
$k_r$ & 0.0008 & -
\\
\hline
$k_p$ & 16 & -
\\
\hline
$k_d$ & 10 & -
\\
\hline
$k_p^q$ & 10 & -
\\
\hline
$K_p^{\omega}$ & \text{diag}([2.5 , 2 , 5]) & -
\\
\hline
$K_d^{\omega}$ & \text{diag}([0.1 , 0.2 , 0.1]) & -
\\
\hline
\end{tabular}
\label{tab1}
\end{center}
\end{table}

\section{Simulation Results}
\label{simulation}

\begin{figure}[!htbp]
\centering
\includegraphics[width=3.4in,height=2.5in]{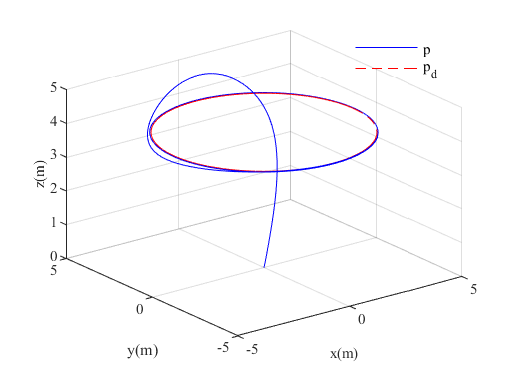}
\caption{The desired and actual position trajectories through the simulations (isometric view).}
\label{3d_position}
\end{figure}

\begin{figure}[htbp]
\centering
\includegraphics[width=2in,height=2in]{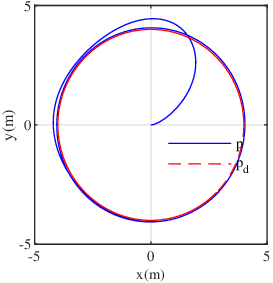}
\caption{The desired and actual position trajectories from the top view ($X_iY_i$-plane).}
\label{2d_position}
\end{figure}

\begin{figure}[htbp]
\captionsetup[subfigure]{labelformat=empty}
\centering
\subfloat[]{\includegraphics[width=3.4in,height=4in]{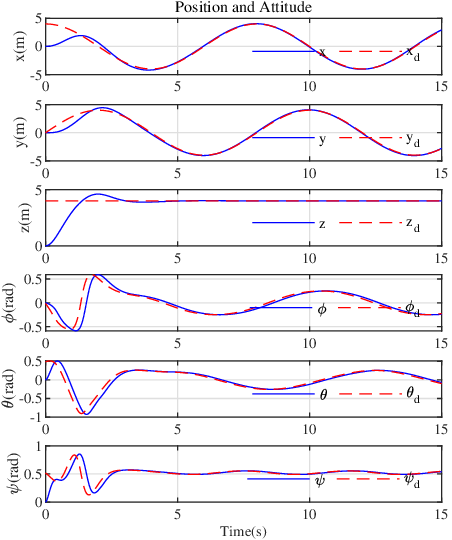}%
}
\hfil
\subfloat[]{\includegraphics[width=3.4in,height=4in]{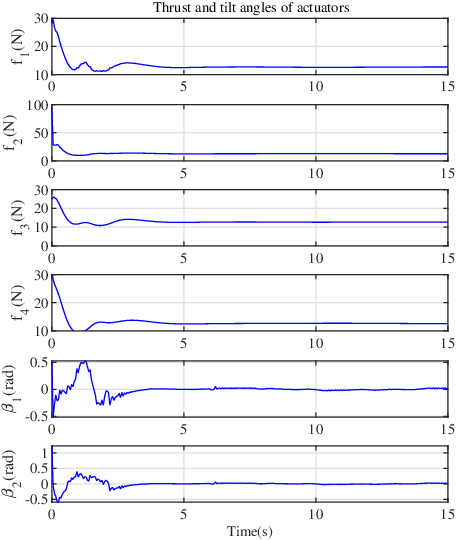}}%
\caption{The desired and actual position and attitude trajectories and the required actuator inputs for proper tracking. }
\label{simulation_results}
\end{figure}

\begin{figure}[htbp]
\centering
\includegraphics[width=3.4in,height=3.2in]{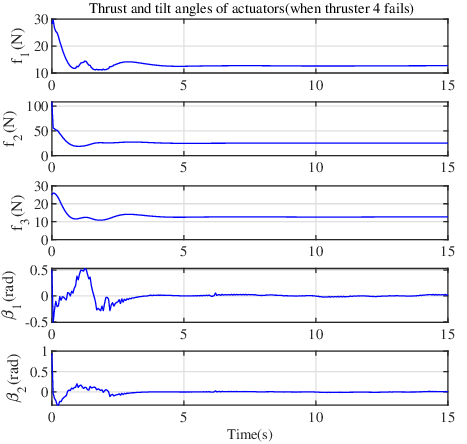}
\caption{The actuator commands required for the safe operation in the case of thruster 4 failure.}
\label{m4}
\end{figure}

\begin{figure}[htbp]
\centering
\includegraphics[width=3.4in,height=2.8in]{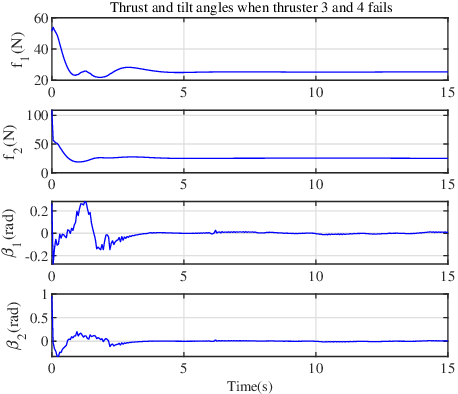}
\caption{The actuator commands required during both the bottom propellers fail case(Bicopter).}
\label{m34}
\end{figure}

The performance of the proposed system dynamics with the proposed tracking controller is tested with simulations using MATLAB. The vehicle parameters and the controller parameters are given in Table \ref{tab1}. The position of the Bi-Quadcopter is intended to follow a circle-shaped trajectory in $XY$ plane by maintaining a constant height with a constant heading angle
\begin{equation*}
\begin{aligned}
    & p_d(t)  = [4\cos(\frac{\pi}{4}t) \  4\sin(\frac{\pi}{4}t) \ 4]^{\top},
    \\
    & X_{bc} = [\cos(\frac{\pi}{6}) \ \sin(\frac{\pi}{6}) \ 0]^{\top}.
\end{aligned}
\end{equation*}

The initial conditions of the Bi-Quadcopter are
\begin{equation*}
\begin{aligned}
   & p(0) = [0 \ 0 \ 0]^{\top} ; \ v(0) = [0 \ 0 \ 0]^{\top},
    \\
   & q(0) = [1 \ 0 \ 0 \ 0]^{\top}; \ \omega(0) = [0 \ 0 \ 0]^{\top}.
\end{aligned}
\end{equation*}
The closed loop system with the dynamics in (\ref{transl}),(\ref{rotat}), and the controllers in (\ref{f_z}),(\ref{omega_d}),(\ref{torque}) is simulated and the position trajectory tracking plots are given in Figure \ref{3d_position} and Figure \ref{2d_position}. For ease of visualization, the quaternion variables are converted to $ZYX$ Euler angles with roll($\phi$), pitch($\theta$), and 
 yaw/heading($\psi$). The position,  attitude, and control trajectories are given in Figure \ref{simulation_results}. It is observed from the plots that the proposed UAV configuration along with the proposed controller is giving good position-tracking performance. 
\subsection{Control re-allocation cases for thruster failures}
The two failure conditions discussed in Section \ref{allocation} are also tested in simulations. 
\subsubsection{Case 1: One thruster failure(thruster 3 or 4)} The actuator inputs $u_1$ are plotted in Figure \ref{m4}. It is observed from Figure \ref{m4} that thruster 2 provides almost half of the total hovering thrust and the other two propellers share the remaining hovering thrust.
\subsubsection{Case 2: Both thrusters fail (thrusters 3 and 4)} The actuator inputs $u_2$ are plotted in Figure \ref{m34}. In this case, the UAV performance is similar to the Bicopter, and the total thrust required is shared by both the top propellers as shown in Figure \ref{m34}. 
\section{Conclusion}
\label{conclusion}
In this paper, a new type of under-actuated tilting UAV with auxiliary thrusters is proposed. The control allocation is discussed to map the actuator inputs to the control wrench. A non-linear position control is implemented and finally, the proposed system with the control is validated through the simulations. The thruster failure analysis and the control re-allocation indicate that the proposed UAV is giving promising performance for real-world applications. 

\bibliographystyle{IEEEtran}
\bibliography{source2}

\end{document}